\documentstyle[epsfig]{article}
\newcommand{\be}{\begin{equation}} \newcommand{\ee}{\end{equation}}
\newcommand{\ba}{\begin{eqnarray}} \newcommand{\ea}{\end{eqnarray}}
\newcommand{\gsim}{
	\renewcommand{\arraystretch}{0.4}\begin{array}{c}>\\ \sim\end{array}}

\begin{document}
\title{\Large \bf
    Transversity distribution function in hard scattering of\\
    polarized protons and antiprotons in the PAX experiment}
\author{A.~V.~Efremov$^a$, K.~Goeke$^b$, P.~Schweitzer$^b$ \\
    \footnotesize\it $^a$ Joint Institute for Nuclear Research,
        Dubna, 141980 Russia\\
    \footnotesize\it $^b$ Institut f\"ur Theoretische Physik II,
        Ruhr-Universit\"at Bochum,  D-44780 Bochum, Germany}
\date{April 2004}
\maketitle

\begin{abstract}
\noindent
    Estimates are given for the double spin asymmetry in lepton-pair
    production from collisions of transversely polarized protons and
    antiprotons for the kinematics of the recently proposed PAX
    experiment at GSI on the basis of predictions for the transversity
    distribution from the chiral quark soliton model.
\end{abstract}

\section{Introduction}
The leading structures of the nucleon in deeply inelastic
scattering processes are described in terms of three twist-2 parton
distribution functions -- the unpolarized $f_1^a(x)$, helicity $g_1^a(x)$,
and transversity $h_1^a(x)$ distribution.
Owing to its chirally odd nature $h_1^a(x)$ escapes measurement in deeply
inelastic scattering experiments which are the main source of information
on the chirally even $f_1^a(x)$ and $g_1^a(x)$.
The transversity distribution function was originally introduced in the
description of the process of dimuon production in high energy collisions
of transversely polarized protons \cite{Ralston:ys}.

Alternative processes have been discussed. Let us mention here the Collins 
effect \cite{Collins:1992kk} which, in principle, allows to access $h_1^a(x)$ 
in connection with a fragmentation function describing a possible spin 
dependence of the fragmentation process, {\sl cf.} also \cite{Mulders:1995dh}
and references therein.
Recent and/or future data from semi-inclusive deeply inelastic scattering 
experiments at HERMES \cite{Airapetian:1999tv}, CLAS \cite{Avakian:2003pk} 
and COMPASS \cite{LeGoff:qn} could be (partly) understood in terms of this 
effect \cite{DeSanctis:2000fh,Ma:2002ns,Efremov:2001cz}.
Other processes to access $h_1^a(x)$ have been suggested as well, {\sl c.f.}
the review \cite{Barone:2001sp}. However, in all these processes $h_1^a(x)$
enters in connection with some unknown fragmentation function. Moreover these
processes involve the introduction of transverse parton momenta, and
for none of them a strict factorization theorem could be formulated
so far. The Drell-Yan process remains up to now the theoretically
cleanest and safest way to access $h_1^a(x)$.

The first attempt to study $h_1^a(x)$ by means of the Drell-Yan process
is planned at RHIC \cite{Bland:2002sd}. The STAR Collaboration has
already delivered data on a single spin asymmetry in the process
$pp^\uparrow\to\pi X$ \cite{Rakness:2002tb} in which $h_1^a(x)$ may be
involved.\footnote{
    The STAR data confirm earlier observations by the FNAL-E704
    collaboration \cite{Adams:1991cs} at substantially higher energies.
    Assuming factorization one finds that $h_1^a(x)$ contributes to this
    process, however, in convolution with the equally unknown Collins
    fragmentation function and in competition with other mechanisms
    \cite{Efremov:1994dg,Anselmino:1994tv}.}
Dedicated estimates, however, indicate that at RHIC the access of
$h_1^a(x)$ by means of the Drell-Yan process is very difficult
\cite{Bunce:2000uv,Bourrely:1994sc}. This is partly due to the kinematics 
of the experiment. The main reason, however, is that the observable double
spin asymmetry $A_{TT}$ is proportional to a product of transversity
quark and antiquark distributions.
The latter are small, even if they were as large as to saturate
the Soffer inequality \cite{Soffer:1994ww} which puts a bound on
$h_1^a(x)$ in terms of the better known $f_1^a(x)$ and $g_1^a(x)$.

This problem can be circumvented by using an antiproton beam
instead of a proton beam. Then $A_{TT}$ is proportional to
a product of transversity quark distributions from the proton
and transversity antiquark distributions from the antiproton
(which are connected by charge conjugation). Thus in this case
$A_{TT}$ is due to valence quark distributions, and one can
expect sizeable counting rates. The challenging program how to
polarize an antiproton beam has been recently suggested in the
{\bf P}olarized {\bf A}ntiproton e{\bf X}periment (PAX) at GSI
\cite{PAX}. The technically realizable polarization of the
antiproton beam of about $(5-10)\%$ and the large counting rates
-- due to the use of antiprotons -- make the program promising.

In this note we shall make quantitative estimates for the Drell-Yan
double spin asymmetry $A_{TT}$ in the kinematics of the PAX experiment.
For that we shall stick to the description of the process at LO QCD.
NLO corrections have been shown to be of order $(10-30)\%$
\cite{Bunce:2000uv,NLO} which is, however, a sufficient accuracy
for our purposes at the present stage.
We also will estimate the recently suggested analog double spin
asymmetry in $J/\Psi$ production \cite{Anselmino-new}.
For the transversity distribution we shall use predictions from the
chiral quark soliton model \cite{Pobylitsa:1996rs,Schweitzer:2001sr}
which have also been used among the first attempts \cite{Efremov:2001cz}
to interpret and/or predict single spin effects at HERMES, CLAS and
COMPASS \cite{Airapetian:1999tv,Avakian:2003pk,LeGoff:qn}.

\section{Lepton pair production in collisions of transversely polarized 
         {\boldmath $p\bar p$}}
The process $p\bar p\to \mu^+\mu^-X$ can be characterized
by the invariants: Mandelstam $s=(p_1+p_2)^2$ and dilepton invariant mass
$Q^2=(k_1+k_2)^2$, where $p_{1/2}$ and $k_{1/2}$ are the momenta of
respectively the incoming proton-antiproton pair and the outgoing lepton
pair, and the rapidity
\be
    y=\frac12\,{\rm ln}\frac{p_1(k_1+k_2)}{p_2(k_1+k_2)}\;.
\ee
Let us denote by $\uparrow(\downarrow)$ the relative orientation
of the transverse polarization of protons and antiprotons.
The double spin asymmetry in Drell-Yan lepton-pair production with
transversely polarized protons and antiprotons is given by
\be\label{Eq:ATT-0}
    \frac{N^{\uparrow\uparrow}-N^{\uparrow\downarrow}}
             {N^{\uparrow\uparrow}+N^{\uparrow\downarrow}}
    = D_P \; f(\theta,\phi) \; A_{TT}(y,Q^2) \;.\ee
The factor $D_P$ takes into account depolarization effects.
(Detector acceptance effects will not be considered.)
The function $f(\theta,\phi)$ is given by
\be\label{Eq:f(angles)}
    f(\theta,\phi) = \frac{\sin^2\theta}{1+\cos^2\theta}\;\cos 2\phi,
\ee
where $\theta$ is the emission angle of one lepton in the dilepton
rest frame and $\phi$ its azimuth angle around the collision axis
counted from the polarization plane of the hadron whose spin is not
flipped in Eq.~(\ref{Eq:ATT-0}). Finally $A_{TT}$ is given by
\be\label{Eq:ATT-1}
    A_{TT}(y,Q^2)=\frac{\sum_a e_a^2 h_1^a(x_1,Q^2) h_1^a(x_2,Q^2)}
                       {\sum_b e_b^2 f_1^b(x_1,Q^2) f_1^b(x_2,Q^2)}\;,
\ee
where the parton momenta $x_{1/2}$ in Eq.~(\ref{Eq:ATT-1})
are fixed in terms of $s$, $Q^2$ and $y$ as
\be\label{Eq:x12}
    x_{1/2} = \sqrt{\frac{Q^2}{s}}\,e^{\pm y}\;.\ee
The sum goes over all quark and antiquark flavours
$a=u,\,\bar u,\,d,\bar d,\,\dots$ etc. In Eq.~(\ref{Eq:ATT-1})
use was made of the charge conjugation invariance which relates
distributions in the nucleon and antinucleon as, e.g.,
\be\label{Eq:cc}
     h_1^{u/p}(x) = h_1^{\bar u/\bar p}(x) \,.
\ee
Distribution functions without explicit indication of the hadron
refer to the proton, i.e. $f_1^u(x)\equiv f_1^{u/p}(x)$, etc.
Eq.~(\ref{Eq:ATT-1}) corresponds to LO QCD. It is modified at NLO
\cite{NLO}.

In the PAX experiment an antiproton beam with energies
in the range $(15-25)\,{\rm GeV}$ could be available, which
yields an $s=(30-50)\,{\rm GeV}^2$ for a fixed proton target.
For this kinematics the "safe region" \cite{McGaughey:1999mq}
for Drell-Yan experiments,
i.e.\  above the region $Q\ge 4\,{\rm GeV}$ dominated by
lepton pairs from leptonic decays of charmed vector mesons,
would mean to probe parton distribution functions in the large
$x$-region $x>0.5$.

The region $1.5\,{\rm GeV} < Q < 3\,{\rm GeV}$, i.e. below the
$J/\Psi$ threshold but well above the region of dileptons from
$\Phi(1020)$-decays (and with sufficiently large $Q^2$ to be in
the hard-scattering regime) would allow to explore the region
$x > 0.2$.
However, in principle one can also address the resonance region itself
-- and benefit from large counting rates \cite{Anselmino-new}.
Whether the "Drell-Yan subprocess" proceeds via
$q\bar q\to\gamma^\ast\to\mu^+\mu^-$ or via
$q\bar q\to J/\Psi \to\mu^+\mu^-$ is irrelevant for $A_{TT}$, since the
unknown $q\bar q J/\Psi$ and $J/\Psi\mu^+\mu^-$--couplings cancel in the
ratio in Eq.~(\ref{Eq:ATT-0}) as argued in Ref.~\cite{Anselmino-new}.

In any case a good understanding of background processes, possible power
corrections and the $K$-factors is required. Low dilepton mass regions
(in nucleon-nucleus collisions) were studied in \cite{Masera:ck}.
Keeping this in mind we shall present below estimates for
$s=45\,{\rm GeV}^2$, and $Q^2=5\,{\rm GeV}^2$, $9\,{\rm GeV}^2$ and
$16\,{\rm GeV}^2$.

\section{Chiral quark-soliton model prediction for {\boldmath $h_1^a(x)$}}
In order to make quantitative estimates for $A_{TT}$ in the PAX kinematics
we will use for the transversity distribution function predictions from
the chiral quark-soliton model. This model was derived from the instanton
model of the QCD vacuum \cite{Diakonov:2002fq} and describes numerous
nucleonic properties without adjustable parameters to within $(10-30)\%$
accuracy \cite{Christov:1995vm}.
The field theoretic nature of the model allows to consistently compute
quark and antiquark distribution functions \cite{Diakonov:1996sr}
which agree with parameterizations \cite{Gluck:1994uf} to within the
same accuracy. This gives us a certain confidence that the model
describes $h_1^a(x)$ with a similar accuracy.

In the chiral-quark soliton model we observe the hierarchy
$h_1^u(x)\gg|h_1^d(x)|\gg|h_1^{\bar u}(x)|$, and an interesting "maximal
sea quark flavour asymmetry" $h_1^{\bar d}(x)\approx -h_1^{\bar u}(x)>0$
\cite{Schweitzer:2001sr}.
In Figure~1a we show the chiral quark soliton model prediction for
$h_1^a(x)$ from Ref.~\cite{Schweitzer:2001sr} LO-evolved from the
low scale of the model of about $\mu_0^2=(0.6\,{\rm GeV})^2$ to the
scale $Q^2=16\,{\rm GeV}^2$.
In order to gain some more intuition on the predictions we compare in Fig.~1b
the dominating distribution function $h_1^u(x)$ from the chiral-quark
soliton model to $f_1^u(x)$ and $g_1^u(x)$ from the parameterizations
of Ref.~\cite{Gluck:1994uf}. It is remarkable that the Soffer inequality
$|h_1^u(x)| \le (f_1^u+g_1^u)(x)/2$ is nearly saturated -- in particular
in the large-$x$ region. (The Soffer bound in Fig.~1b is constructed from
$f_1^u(x)$ and $g_1^u(x)$ taken at $Q^2=16\,{\rm GeV}^2$ from
\cite{Gluck:1994uf}.)

For the unpolarized distribution function $f_1^a(x)$ we
use the LO parameterization from Ref.~\cite{Gluck:1994uf}.

\section{Double spin asymmetry {\boldmath $A_{TT}$} at PAX}
The estimates for the double spin asymmetry $A_{TT}$ as defined in
Eq.~(\ref{Eq:ATT-1}) for the PAX kinematics on the basis of the
ingredients discussed above is shown in Fig.~2a.
The explorable rapidity range shrinks with increasing dilepton
mass $Q^2$. Since $s=x_1x_2Q^2$, for $s=45\,{\rm GeV}^2$ and
$Q^2=5\,{\rm GeV}^2$ ($16\,{\rm GeV}^2$) one probes parton momenta
$x>0.3$ ($x>0.5$).
The asymmetry $A_{TT}$ grows with increasing $Q^2$ where larger parton
momenta $x$ are involved, since $h_1^u(x)$ is larger with respect to
$f_1^u(x)$ in the large $x$-region, {\sl cf.} Fig.~1b. The magnitude of 
$A_{TT}$ can roughly be estimated by noting that at $Q^2=5\,{\rm GeV}^2$ 
in the model with a good accuracy $xh_1^u(x)\approx 4.0 \, x(1-x)^3$ 
for all $x$, while somehow more roughly $xf_1^u(x)\approx 5.5\, x(1-x)^3$ 
for $x\gsim 0.5$.
This yields $A_{TT}\approx (4.0/5.5)^2\approx 0.5$ considering 
$u$-quark dominance.

    \begin{figure}
    \begin{tabular}{cc}
    \epsfig{figure=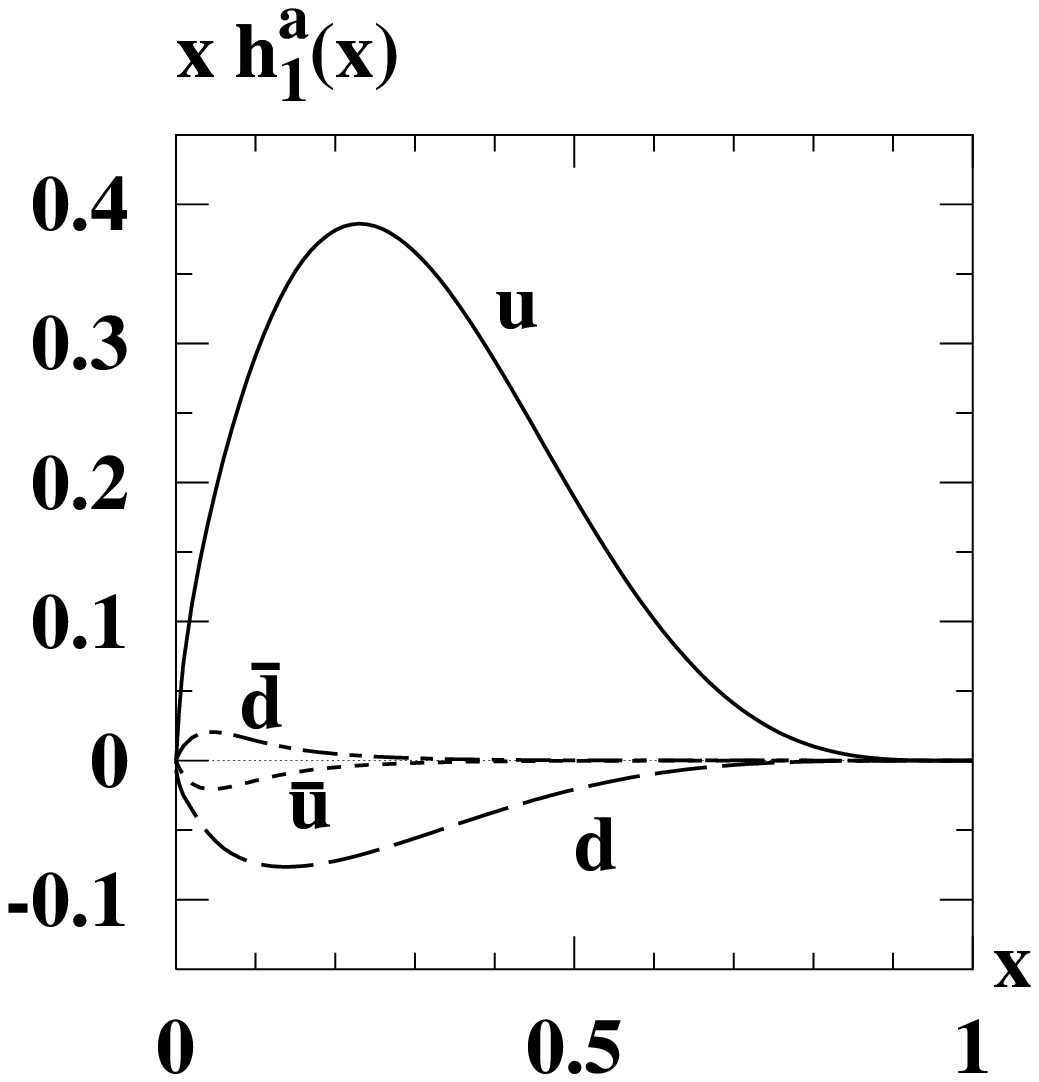,       width=6.5cm,height=6cm} &
    \epsfig{figure=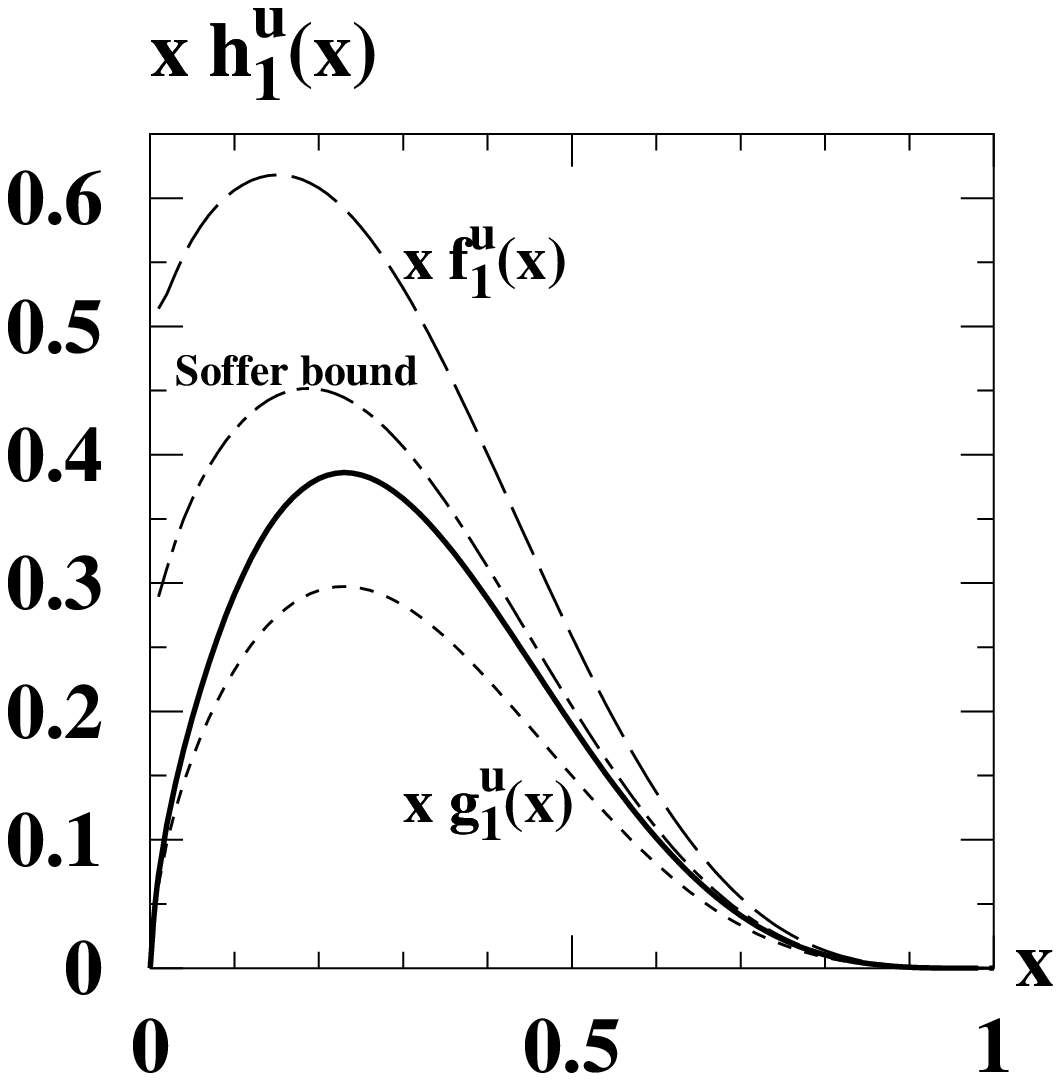,width=6.5cm,height=6cm} \cr
    {\bf a} & {\bf b}
    \end{tabular}
    \caption{
    {\bf a.}
    The transversity distribution function $h_1^a(x)$ vs.\  $x$
    from the chiral quark soliton model \cite{Schweitzer:2001sr}.
    {\bf b.}
    Comparison of $h_1^u(x)$ from the chiral quark soliton model (solid)
    to $f_1^u(x)$ (dashed) and $g_1^u(x)$ (dotted) and the Soffer bound
    $(f_1^u+g_1^u)(x)/2$ (dashed-dotted line) with the parameterizations
    of \cite{Gluck:1994uf}.
    All curves in Figures~1a and 1b are multiplied by $x$ and are LO
    evolved to a scale of $Q^2=16\,{\rm GeV}^2$.}
    \begin{tabular}{cc}
    \epsfig{figure=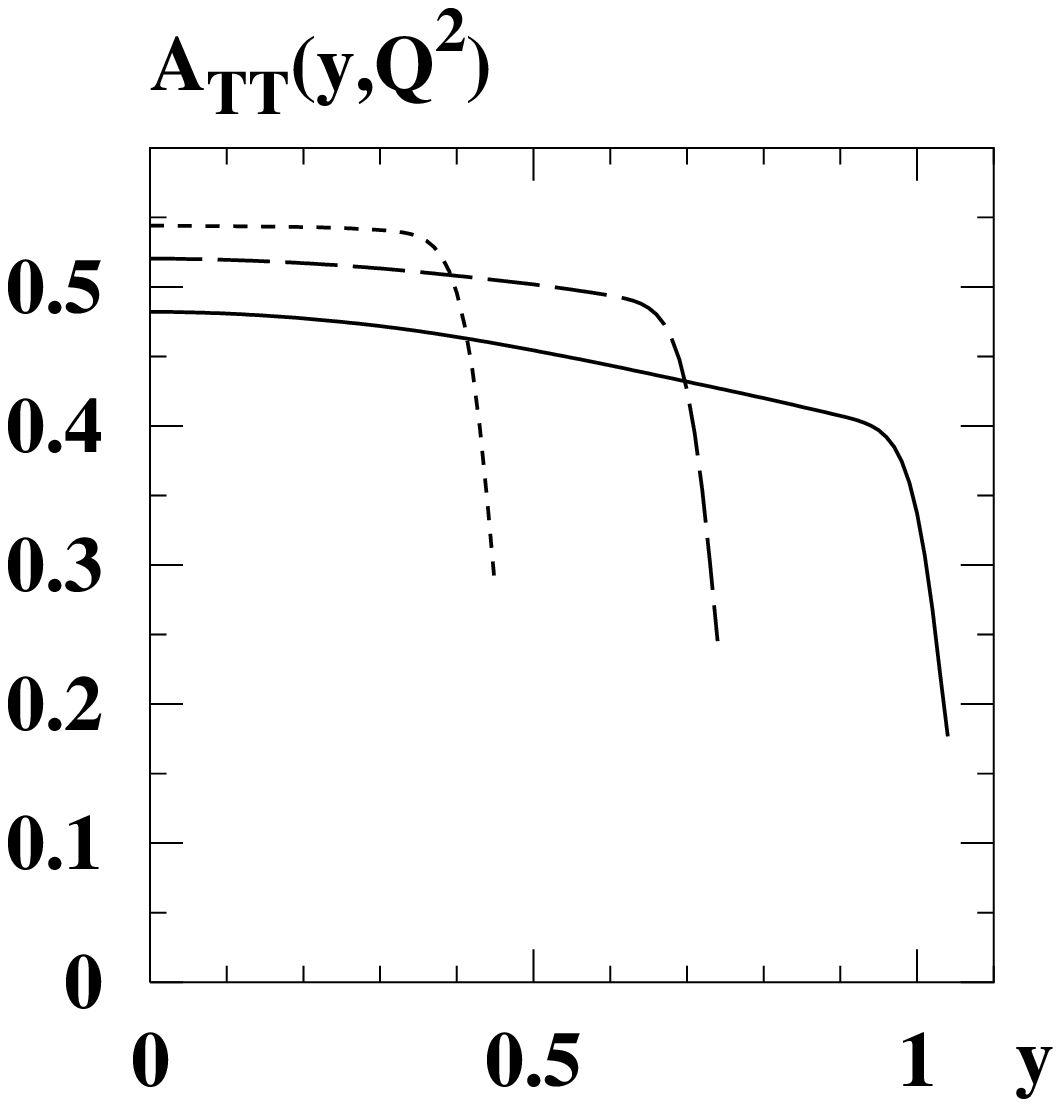,width=6.5cm,height=6cm} &
    \epsfig{figure=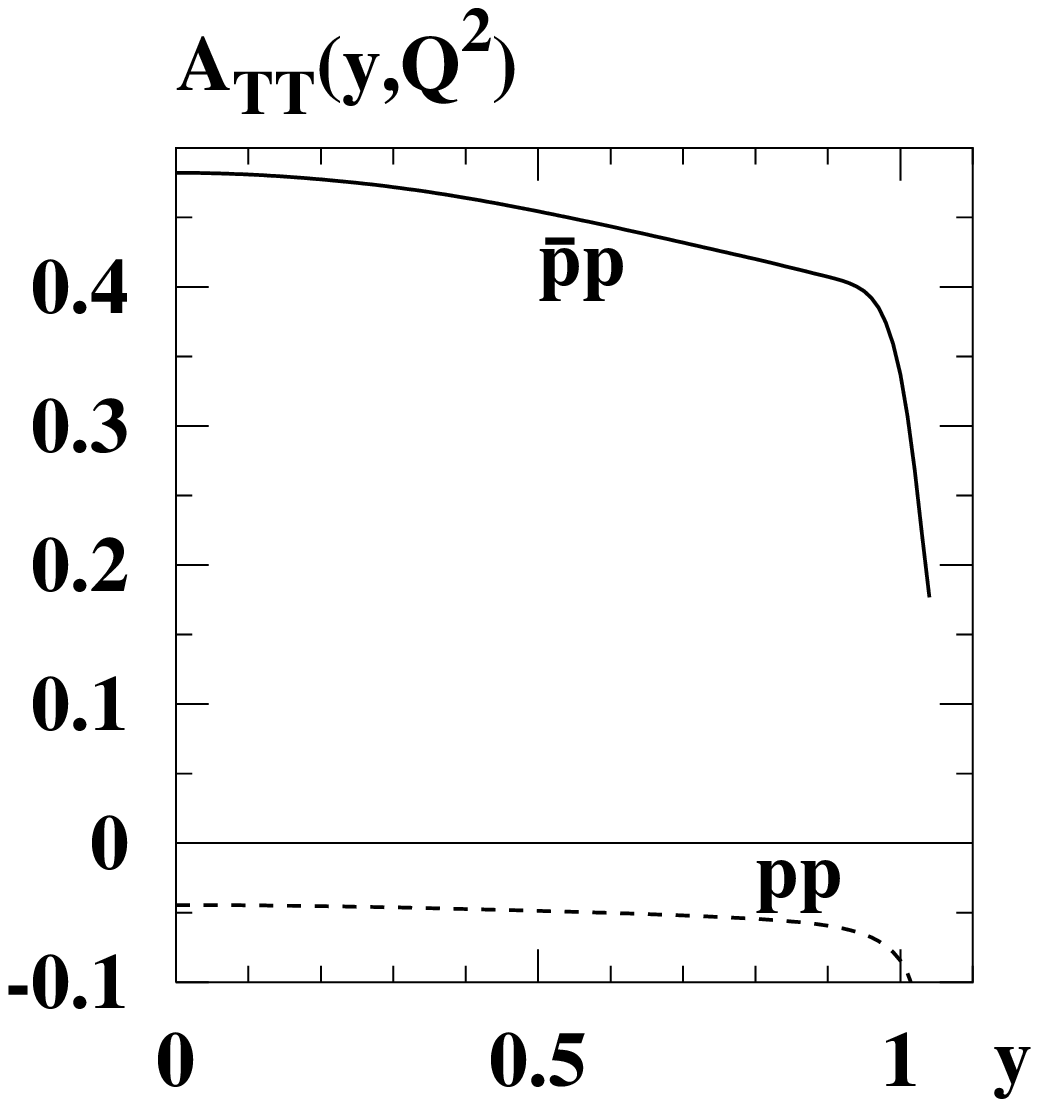, width=6.5cm,height=6cm} \cr
    {\bf a} & {\bf b}
    \end{tabular}
    \caption{
    {\bf a.}
    The asymmetry $A_{TT}(y,M^2)$, {\sl cf.} Eq.~(\ref{Eq:ATT-1}),
    as function of the rapidity $y$ for $Q^2=5\,{\rm GeV}^2$ (solid)
    and $9\,{\rm GeV}^2$ (dashed) and $16\,{\rm GeV}^2$ (dotted line)
    for $s=45\,{\rm GeV}^2$.
    {\bf b.} Comparison of $A_{TT}(y,M^2)$ from proton-antiproton
    (solid) and proton-proton (dotted line) collisions at PAX for
    $Q^2=5\,{\rm GeV}^2$ and $s=45\,{\rm GeV}^2$.}
    \end{figure}

The advantage of using antiprotons is evident from Fig.~2b.
The corresponding asymmetry from proton-proton collisions is an order of
magnitude smaller (this observation holds also in the kinematics of RHIC
\cite{Schweitzer:2001sr}).
At first glance this advantage seems
to be compensated by the polarization factor in Eq.~(\ref{Eq:ATT-0}).
$D_P$ is basically the product of the antiproton beam polarization of
$(5-10)\%$ and the proton target polarization of $90\%$, i.e.\ at
PAX $D_P\approx 0.05$. E.g., at RHIC the polarization of each proton
beam could reach $70\%$ yielding $D_P\approx 0.5$. However, thanks to the
use of antiprotons cross sections and counting rates are more sizeable.

A precise measurement of $A_{TT}$ in the region
$Q>4\,{\rm GeV}$ is very difficult, however, in the dilepton mass region
below the $J/\Psi$ threshold \cite{PAX} and in the resonance region
\cite{Anselmino-new} $A_{TT}$ could be measured with sufficient accuracy
in the PAX experiment.

    \begin{figure}
    \begin{tabular}{cc}
    \epsfig{figure=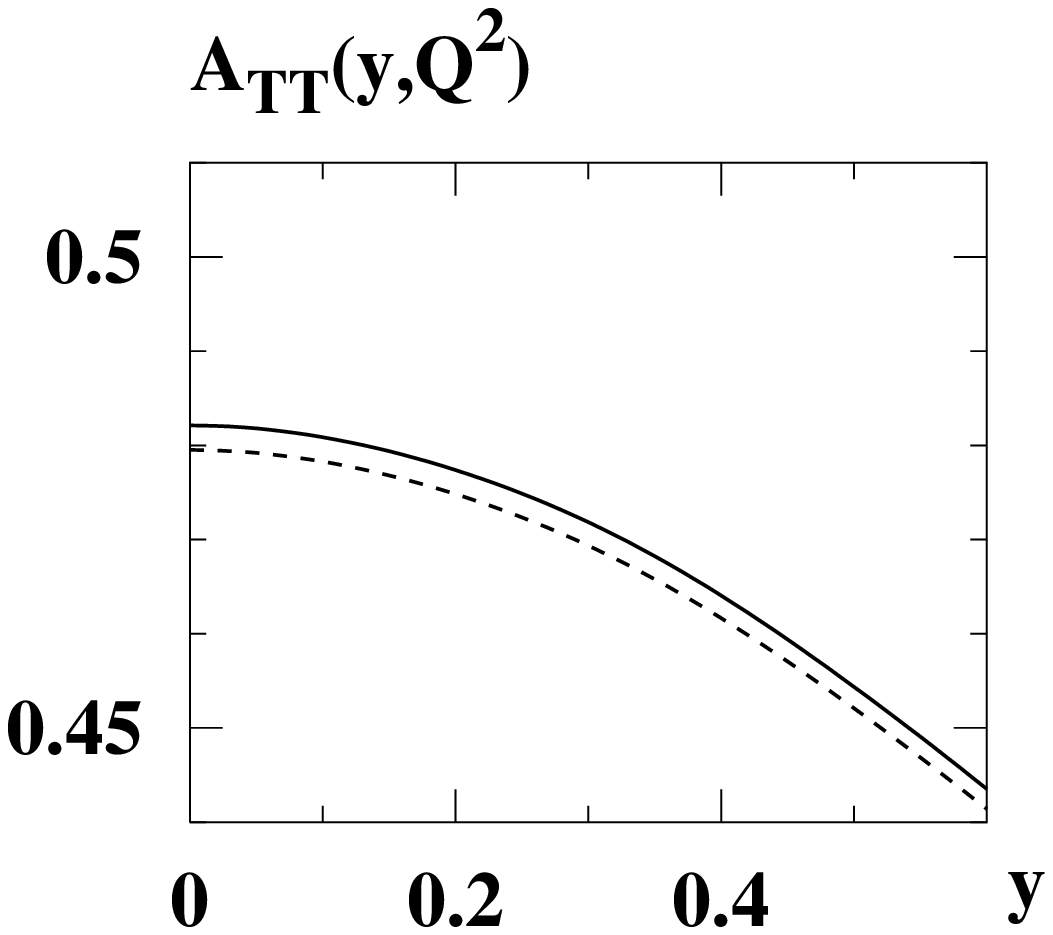,width=6.5cm,height=5.5cm} &
    $\phantom{X}$

    \vspace{-4cm}
    {\begin{minipage}[b]{8cm}
    Figure 3: The Drell-Yan double spin asymmetry $A_{TT}$ at PAX
    for $s=45\,{\rm GeV}^2$ and $Q^2=5\,{\rm GeV}^2$. Solid line:
    The full result. Dashed line: The "transversity $u$-quark
    approximation"; only $h_1^u(x)$ is considered in the numerator
    of $A_{TT}$ in Eq.~(\ref{Eq:ATT-1}).
    $\phantom{XXXXXXXXXXXXXXXXXXXXXXXXX}$
    $\phantom{XXXXXXXXXXXXXXXXXXXXXXXXX}$
    $\phantom{XXXXXXXXXXXXXXXXXXXXXXXXX}$
    \end{minipage}}
    \vspace{4cm}
    \end{tabular}
\end{figure}

What could one learn from a measurement of the Drell-Yan double
spin asymmetry $A_{TT}$ in proton-antiproton collisions at PAX?
The PAX experiment is sensitive in particular to $h_1^u(x)$.
This is demonstrated by Fig.~3
where $A_{TT}$ is compared to what one would obtain in a
"transversity $u$-quark-only approximation", i.e. by replacing
$\sum_a e_a^2 h_1^a(x_1)h_1^a(x_2)\to\frac49h_1^u(x_1)h_1^u(x_2)$ in the
numerator of $A_{TT}$ in Eq.~(\ref{Eq:ATT-1}). Clearly, with good
accuracy the result can be interpreted as being due to $h_1^u(x)$ only.

In the mid-rapidity region $y\approx 0$ the asymmetry
$A_{TT}\propto [h_1^u(x)]^2$ at $x\approx\sqrt{Q^2/s}$. A precise
measurement would allow to discriminate between different models
for $h_1^a(x)$. E.g., on the basis of the non-relativistic quark model
motivated popular guess $h_1^a(x)\approx g_1^a(x)$ (at some unspecified
low scale) one would expect an $A_{TT}$ of about $30\%$
\cite{Anselmino-new} to be contrasted with the chiral quark soliton
model estimate of about $50\%$, {\sl cf.} Fig.~2a.

\section{Summary}
To summarize, in the recently proposed experiment PAX at GSI
one could access the $u$-quark transversity distribution function
in the valence $x$-region $x>0.2$ by means of the double-spin
asymmetry in Drell-Yan lepton pair production from collisions of
transversely polarized protons and antiprotons \cite{PAX}.
A leading order QCD estimate yields at dilepton invariant masses below
the threshold of $J/\Psi$, but well above the background from decays of
$\Phi(1020)$ and other resonances, sizeable spin asymmetries
$A_{TT}\approx (40-50)\%$ on the basis of predictions for $h_1^a(x)$
from the chiral-quark soliton model \cite{Schweitzer:2001sr}.
At next-to-leading order in QCD one can expect corrections to this
result which reduce somehow the asymmetry \cite{NLO}.
Similarly large asymmetries can be also expected in the recently suggested
process of lepton pair production via $J/\Psi$ production \cite{Anselmino-new}.
In order to unambiguously interpret the result it is, however, necessary
to understand well -- both phenomenologically and theoretically -- background
processes, possible power corrections and $K$-factors in the dilepton mass
region $Q<4\,{\rm GeV}$.

\paragraph{Acknowledgements.}
We thank the PAX collaboration for attracting our attention to the possibility
of accessing $h_1^a(x)$ by means of polarized proton-antiproton scattering, and
are grateful to Andreas Metz for encouragement and fruitful discussions.
A.~E.\ is partially supported by INTAS grant 00/587,
RFBR grant 03-02-16816 and DFG-RFBR 03-02-04022.
This work is partially supported by Verbundforschung of BMBF.

\end{document}